\documentclass[10pt,letterpaper,twocolumn]{article} 

\usepackage{ol2}
\usepackage[draft]{hyperref}
\usepackage{amsmath}
\begin{document}

\twocolumn[ 

\title{Correlated imaging through atmospheric turbulence}


\author{Pengli Zhang, Wenlin Gong, Xia Shen and Shensheng Han$^{*}$}

\address{Key Laboratory for Quantum Optics and Center for Cold Atom Physics,
Shanghai Institute of Optics and Fine Mechanics, Chinese Academy of
Sciences, Shanghai 201800, China }
\address{$^*$Corresponding author: sshan@mail.shcnc.ac.cn}

\begin{abstract} Correlated imaging through atmospheric
turbulence is studied, and the analytical expressions describing
turbulence effects on image resolution are derived. Compared with
direct imaging, correlated imaging can reduce the influence of
turbulence to a certain extent and reconstruct high-resolution
images. The result is backed up by numerical simulations, in which
turbulence-induced phase perturbations are simulated by random phase
screens inserting propagation paths.
\end{abstract}


 ] 
As correlated imaging develops well in recent years
\cite{Bromberg,Gong,Shapiro,Chan,Cheng}, more attention has been
focused on how to apply this technique to practical applications to
overcome the limits in conventional optical systems. For an imaging
system which must look through the atmosphere, turbulence-induced
wavefront variations distort the point spread function (PSF) of the
system from its ideal diffraction-limited shape, which leads to the
the degradation of image resolution \cite{Roggemann}. To mitigate
turbulence effects, a number of methods, such as speckle imaging and
adaptive optics techniques \cite{Roggemann}, have been proposed and
applied in optical astronomy. Nonetheless, each of these techniques
has its own set of performance limits, hardware and software
requirements. New approaches to the problem of reducing these
effects are still of much interest. Here we investigate the
performance of correlated imaging through atmospheric turbulence and
find that the influence of turbulence can be weakened by the
second-order intensity correlation.

\begin{figure}[htb]
\centerline{\includegraphics[width=8.3cm]{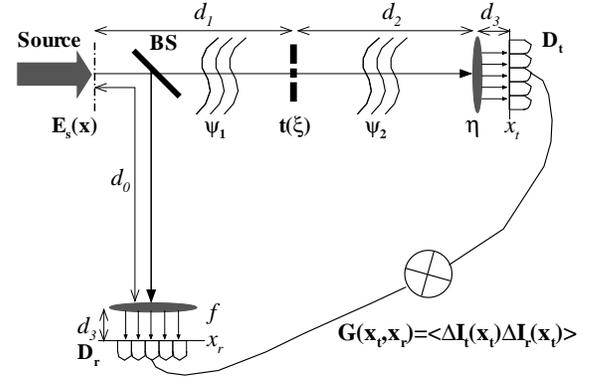}}
 \caption{Schematic of correlated imaging through atmospheric
turbulence.} \label{fig1}
\end{figure}

   A schematic of correlated imaging through the atmosphere is
depicted in Fig. \ref{fig1}. The beam splitter (BS) divides thermal
light into two beams propagating through two distinct optical paths.
One is test arm which includes an unknown object and a telescope
setup consisting of a lens with focal length $f$ and a detector
$D_{t}$. The object is located at a distance $d_{1}$ from the source
as well as $d_{2}$ to the telescope setup. The other is the
reference arm where another telescope setup consisting of a lens and
a detector $D_{r}$ is placed at $d_{0}=d_{1}+d_{2}$ from the source.
For remote sensing (i.e., $d_{1},d_{2}\gg f$), the detector $D_{t}$
(or $D_{r}$) generally lies close to the back focal plane of the
lens (i.e., $d_{3}\approx f$). The test arm is imbedded in the
atmosphere, and turbulence-induced wavefront fluctuations in
propagation paths $d_{1}$ and $d_{2}$ are represented by $\Psi_{1}$
and $\Psi_{2}$, respectively. While the reference arm is said to be
a free-space propagation through the distance $d_{0}$ by assuming
that there exists no turbulence. The assumption is based on the fact
that the optical field in the reference arm is totally predictable
if the field distribution of the source is well known
\cite{Bromberg,Shapiro}.

In the test arm, the field $E_{t}(x_{t})$ in the detector $D_{t}$
can be given by
\begin{eqnarray}\label{Eq1}
E_{t}(x_{t})=\iint dxd\xi E_{s}(x) h_{1}(\xi, x) t(\xi)
h_{2}(x_{t},\xi),
\end{eqnarray}
where $E_{s}(x)$ corresponds to the source field, and $t(\xi)$
denotes the transmission function of the object. $h_{1}(\xi, x)$,
$h_{2}(x_{t},\xi)$ are the impulse response functions from the
source to the object and from the object to the detector $D_{t}$,
respectively.

Furthermore, according to the extended Huygens-Fresnel integral
\cite{Ricklin}, $h_{1}(\xi, x)$ and $h_{2}(x_{t},\xi)$ have the
forms \addtocounter{equation}{1}
\begin{align}\label{Eq2a}
&h_{1}(\xi,x)=\frac{1}{\sqrt{j\lambda
d_{1}}}e^{\frac{jk}{2d_{1}}(x-\xi)^{2}+\Psi_{1}(x,\xi)}
,\tag{\theequation a}\\\label{Eq2b}
&h_{2}(x_{t},\xi)=\frac{1}{j\lambda\sqrt{d_{2}d_{3}}}\int d\eta e^{-
\frac{jk}{d_{2}}(\xi-x_{t}/M)\eta+\Psi_{2}(\xi,\eta)},\tag{\theequation
b}
\end{align}
where $k=2\pi/\lambda$ is the wave number with $\lambda$ being the
wavelength, and $M=-d_{3}/d_{2}$ is the magnification of the
telescope setup. $\Psi_{1}(x,\xi)$ and $\Psi_{2}(\xi,\eta)$ account
for the random parts (due to atmospheric turbulence) of the complex
phases of the fields in the propagation paths $d_{1}$ and $d_{2}$,
respectively.

The field $E_{r}(x_{r})$ in the detector $D_{r}$ is connected to the
source field $E_{s}(x)$ by the Fresnel diffraction integral
\begin{eqnarray}\label{Eq3}
E_{r}(x_{r})=\frac{1}{\sqrt{j\lambda d_{1}|M|}}\int dx
E_{s}(x)e^{\frac{jk}{2d_{1}}(x-x_{r}/M)^2}.
\end{eqnarray}
It's worth pointing out that the apertures of the lenses are
regarded as large enough, and the diffraction limit of the lenses
has been neglected here.

 Performing the intensity correlation measurement between the test arm and the reference arm, we
 get
\begin{eqnarray}\label{Eq4}
G(x_{t},x_{r})&=&\langle I_{t}(x_{t})I_{r}(x_{r})\rangle-\langle I_{t}(x_{t})\rangle\langle I_{r}(x_{r})\rangle\nonumber\\
&=&c_{0}\int dx dx'dx''dx'''
d\xi d\xi' \langle E_{s}(x)E^{\ast}_{s}(x''')\rangle\nonumber\\
& &\times\langle E^{\ast}_{s}(x')E_{s}(x'')\rangle \langle h_{1}(\xi,x)h^{\ast}_{1}(\xi',x')\rangle\nonumber\\
& & \times
\langle h_{2}(x_{t},\xi)h^{\ast}_{2}(x_{t},\xi')\rangle t(\xi)t^{\ast}(\xi')\nonumber\\
& & \times e^{\frac{jk}{2d_{1}}[(x''-x_{r}/M)^2-(x'''-x_{r}/M)^2]},
\end{eqnarray}
where $c_{0}$ is a constant $(\lambda^{3}d_{1}d_{2}d_{3}|M|)^{-1}$,
and $I_{t}(x_{t}),\ I_{r}(x_{r})$ represent the intensity
distributions in $D_{t}$ and $D_{r}$, respectively. Here, we have
supposed that the thermal field, and the two turbulent regions are
statistically independent of each other.

   If the source is fully spatially incoherent and its intensity distribution is of
the Gaussian type, the first-order correlation function of the
source has the form
\begin{eqnarray}\label{Eq5}
\langle
E_{s}(x)E^{\ast}_{s}(x')\rangle=I_{0}e^{-\frac{x^{2}+x'^{2}}{r_{e}^{2}}}\delta(x-x'),
\end{eqnarray}
where $I_{0}$ denotes the mean intensity at the center of the
source, and $r_{e}$ is the $1/e^{2}$ intensity radius. With the help
of Eqs. (\ref{Eq2a}), (\ref{Eq2b}), and (\ref{Eq5}), Eq. (\ref{Eq4})
can be rewritten as
\begin{eqnarray}\label{Eq6}
G(x_{t},x_{r})&=&I^2_0\int dx dx'd\eta d\eta'd\xi d\xi'
t(\xi)t^{\ast}(\xi')
\nonumber\\
& &\times e^{-\frac{2(x^{2}+x'^{2})}{r_{e}^{2}}}
e^{\frac{jk}{2d_{1}}[(x'-x_{r}/M)^2-(x-x_{r}/M)^2]}\nonumber\\
& &\times e^{\frac{jk}{2d_{1}}[(x-\xi)^2-(x'-\xi')^2]}\langle e^{\Psi_1(x,\xi)+\Psi^{\ast}_1(x',\xi')}\rangle\nonumber\\
& &\times
e^{\frac{jk}{d_{2}}[(\xi-x_{t}/M)\eta-(\xi'-x_{t}/M)\eta']}\nonumber\\
& &\times \langle
e^{\Psi_2(\xi,\eta)+\Psi^{\ast}_2(\xi',\eta')}\rangle.
\end{eqnarray}
The ensemble average of phase variations arising from turbulence can
be approximated by \cite{Ricklin}
\begin{eqnarray}\label{Eq7}
& &\langle e^{\Psi_{i}(x,\xi)+\Psi^{\ast}_{i}(x',\xi')}\rangle\nonumber\\
&\cong&
e^{-\frac{1}{\rho^2_i}[(x-x')^2+(x-x')(\xi-\xi')+(\xi-\xi')^2]},
\end{eqnarray}
where $\rho_i=(0.545C^{2(i)}_nk^2d_i)^{-3/5}$ ($i=1,2$) is the
coherence length of a spherical wave propagating in the turbulent
medium and $C^{2(i)}_n$ corresponds to the refractive-index
structure constants describing the strength of atmospheric
turbulence in the propagation path $d_{i}$. It's worth emphasizing
that we have adopted a quadratic approximation of the Rytov's phase
structure function in Eq. (\ref{Eq7}) to obtain  the analytical
formula, and this approximation has been used widely in literatures
\cite{Ricklin,Cheng}.

Substituting Eq. (\ref{Eq7}) to Eq. (\ref{Eq6}) and integrating
 over $\eta, \eta',x,x'$, we have
\begin{eqnarray}\label{Eq8}
G(x_{t},x_{r})&=&\frac{\sqrt{\pi}I^{2}_{0}c_{0}}{\sqrt{\alpha\beta_2(\alpha+2\beta_1)}}
\int d\xi
|t(\xi)|^2\nonumber\\
& &\times
e^{-\frac{2A^2}{\alpha+2\beta_1}(\xi-x_r/M)^2}e^{-\frac{B^2}{\beta_2}(\xi-x_t/M)^2},
\end{eqnarray}
where $A=k/2d_{1}$, $B=k/2d_{2}$, $\alpha=r_{e}^{-2}/2$,
 $\beta_{i}=\rho_{i}^{-2}$ .

   By making $x_{r}=x_{t}$ in Eq. (\ref{Eq8}), we carry out a special
point-to-point intensity correlation \cite{Zhang} and obtain the
 PSF of the correlated imaging system
\begin{eqnarray}\label{Eq9}
h_g(x_{r},\xi)=e^{-\frac{2A^2}{\alpha+2\beta_1}(\xi-x_{r}/M)^2}e^{-\frac{B^2}{\beta_2}(\xi-x_{r}/M)^2}.
\end{eqnarray}

   For the sake of comparison, we also present the intensity distribution in
 $D_{t}$,
 \begin{eqnarray}\label{Eq10}
I_{t}(x_{t})=\frac{\sqrt{\pi}I_{0}c_{0}}{\sqrt{\alpha\beta_{2}}}\int
d\xi |t(\xi)|^{2}e^{-\frac{B^2}{\beta_2}(\xi-x_t/M)^2},
\end{eqnarray}
 and the PSF of the test arm
\begin{eqnarray}\label{Eq11}
h_{t}(x_{t},\xi)=e^{-\frac{B^2}{\beta_2}(\xi-x_t/M)^2}.
\end{eqnarray}

From Eqs. (\ref{Eq9}) and (\ref{Eq11}), we can see that the full
widths at half maximum (FWHM) of $h_{g}$ and $h_{t}$ both broaden
with the increase of $\beta_{i}$ (apart from the influence of the
size of the source), which indicates that the resolution, whether
for correlated imaging or direct imaging, is degraded by atmospheric
turbulence. Additionally, and most importantly, $h_{g}$ has a
narrower FWHM compared to $h_{t}$ , which means that correlated
imaging is helpful to reduce turbulent effects and achieve
high-resolution images.

 In simulations, we consider correlated imaging through
horizontal paths in the atmosphere, and thus $C_n^2$ can be regarded
as constant in the whole turbulent regions. The numerical model of
light propagation in turbulence has been developed well
\cite{Belmonte,Knepp}. The spatial power spectral density of the
index of refraction fluctuations can be described by the Von Karman
spectrum \cite{Belmonte},
\begin{eqnarray}\label{Eq12}
\Phi_n(K,z)=0.033C_n^2(z)(K^2+L_0^{-2})^{-11/6}e^{-(Kl_{0}/2\pi)^{2}},
\end{eqnarray}
where $K^{2}=K^{2}_{x}+K^{2}_{y}+K^{2}_{y}$, $z$ is the propagation
distance from the source, $L_{0}$ and $l_{0}$ represent the outer
scale and inner scale of the turbulence, respectively. By using the
spectrum in Eq. (\ref{Eq12}) to filter a complex Gaussian
pseudorandom field and inverse transforming the result, one obtains
a two-dimensional phase screen which has the same statistics as the
turbulence-induced phase variations \cite{Belmonte}. For long
atmospheric paths, the multiple phase-screen model \cite{Knepp} has
been used in simulations. The turbulent region with the propagation
length $d_{i}$ is broken into a number of layers with a thickness
$\Delta z$. Phase fluctuations in each layer are represented by a
phase screen inserting in the middle of the layer. The effect of
field propagation through these continuous layers can be calculated
separately and then combined to characterize propagation through the
entire turbulent region, provided the index of refraction
fluctuations for each layer are statistically independent
\cite{Roggemann}.

\begin{figure}[htb]
\centerline{\includegraphics[width=8.3cm]{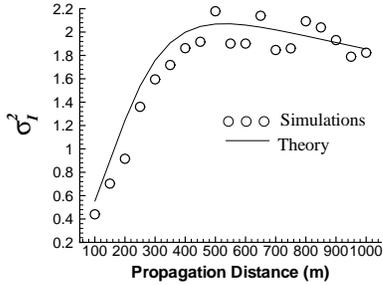}}
 \caption{Simulated ( open circles) and theoretical (solid line)
 on-axis irradiance variance versus the propagation distance.
 The outer scale and inner scale of turbulence are $L_{0}=3$ m
 and $l_{0}=1$ cm, respectively.} \label{fig2}
\end{figure}
  First of all, to verify the computer programm, we investigate the
behavior of a Gaussian beam (waist radius $w_{0}=7$ cm and
wavelength $\lambda=2\ \mu$m) traveling through the atmosphere with
a strong turbulence level ($C_{n}^{2}$=$10^{-12}\textrm{m}^{-2/3}$).
The thickness of each layer is $\Delta z=50$ m. The on-axis
normalized intensity  variance, defined as $\sigma_{I}^{2}=\langle
I^{2}\rangle/\langle I\rangle^{2}-1$ \cite{Belmonte}, is plotted as
a function of the propagation distance in Fig. \ref{fig2}. The good
coincidence between the simulated data (open circles) and the
theoretical result (solid line) predicted by \cite{Andrews} proves
the validity of the programm.

  After the validation, we apply the programm to simulate the
correlated imaging system shown in Fig. \ref{fig1}. The thermal
source ($\lambda=0.532\ \mu$m and diameter $D=2r_{e}= 5$ cm) was
described by a grid of $512\times512$ with a sample spacing $\Delta
x=\Delta y=5$ mm. The distances were set as $d_{1}=d_{2}=10$ km and
the focal length $f=1$ m. The turbulence regions in the paths $d_{i}
(i=1,2)$ were divided into 20 layers with a thickness $\Delta z=500$
m, respectively. The turbulent parameters were assumed constant at
the outer scale $L_{0} = 100$ m and the inner scale $l_{0}=5$ mm. By
averaging over $10^4$ samples, simulated results [{see Fig. (3)]
clearly show the image resolution decrease with the increase of the
turbulent strength, which accords with the analytical calculation
from Eq. (\ref{Eq8}).
\begin{figure}[htb]
\centerline{\includegraphics[width=8.3cm]{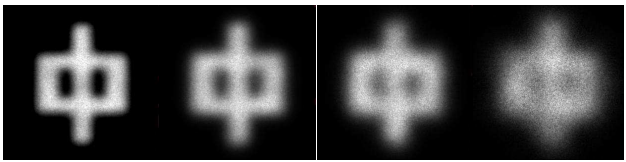}}
 \caption{The reconstructed images (from left to right) via the
 correlation in the atmosphere with turbulent levels
 $C_{n}^{2}=10^{-16}\textrm{m}^{-2/3},\
 2.5\times10^{-16}\textrm{m}^{-2/3},\
 5\times10^{-16}\textrm{m}^{-2/3}$, and $10^{-15}\textrm{m}^{-2/3}$, respectively.
} \label{fig4}
\end{figure}

\begin{figure}[htb]
\centerline{\includegraphics[width=8.3cm]{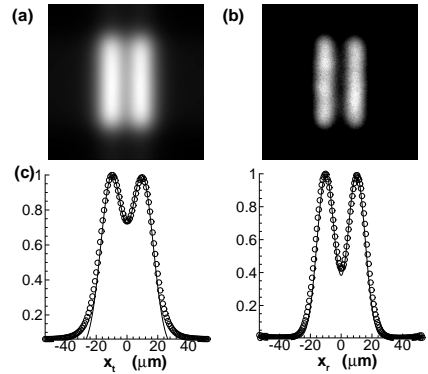}}
 \caption{The acquired images of the double slit in the atmosphere with
  turbulent level $C_{n}^{2}=10^{-15}\textrm{m}^{-2/3}$ .
 (a) was obtained by the test arm, and (b) was extracted from the
 correlation. The
 normalized horizontal sections of the images are plotted in (c),
 where open circles correspond to the simulated data and solid lines show
 the theoretical predictions from Eqs. (\ref{Eq8}) and (\ref{Eq10}), respectively.
} \label{fig3}
\end{figure}

   To compare direct imaging and correlated imaging, a simple double slit
(slit width 10 cm and center-to-center separation 20 cm)
 was used. After statistics over $10^4$ samples, we obtained a blurred image
detected by the test arm directly [see Fig. \ref{fig3}(a)] and a
clear image reconstructed through the correlation [see Fig.
\ref{fig3}(b)]. This confirms the analytical result that ghost
imaging could reduce turbulent effects and improve resolution.

   In summary, by taking advantage of  the extended Huygens-Fresnel
integral, we have presented the  theoretical expressions that
describes how atmospheric turbulence corrupts the image resolution.
Meanwhile, the analytical calculations and the numerical simulations
have demonstrated that correlated imaging can provide imaging
performance superior to direct imaging through the atmosphere. As an
unique image-formed method, correlated imaging can be effectively
combined with conventional phase compensating techniques (e.g.,
adaptive optics) to further eliminate turbulent effects.

   This work is supported by the Hi-Tech Research and development Programm
of China (Grant No. 2006AA12Z115), Shanghai Fundamental Research
Project (Grant No. 09JC1415000), and the National Natural Science
Foundation of China (Grant No. 6087709).

\end{document}